\documentclass{aa}
 \usepackage{graphics}
 
\begin{document}

  \title{The {\em XMM-Newton} Serendipitous Survey
\thanks{Based on observations obtained with {\em XMM-Newton}, an ESA science 
            mission with instruments and contributions directly funded by 
            ESA Member States and the USA (NASA)}
}
 \subtitle{I. The role of {\em XMM-Newton} Survey Science Centre}
 
\author{M.~G.~Watson\inst{1}
\and J-L.~Augu\`eres\inst{2}
\and J.~Ballet\inst{2}
\and X.~Barcons\inst{3}
\and D.~Barret\inst{4}
\and M.~Boer \inst{4}
\and Th.~Boller\inst{5}
\and G.~E.~Bromage\inst{6}
\and H.~Brunner\inst{7}
\and F.~J.~Carrera\inst{8}$^,$\inst{3}
\and M.~S.~Cropper\inst{8}
\and M.~Denby \inst{1}
\and M.~Ehle \inst{5}$^,$\inst{17}
\and M.~Elvis \inst{9}
\and A.~C.~Fabian\inst{10}
\and M.~J.~Freyberg\inst{5}
\and P.~Guillout\inst{11}
\and J-M.~Hameury\inst{11}
\and G.~Hasinger\inst{7}
\and D.~A.~Hinshaw\inst{12}$^,$\inst{1}
\and T.~Maccacaro\inst{13}
\and K.~O.~Mason \inst{8}
\and R.~G.~McMahon\inst{10}
\and L.~Michel\inst{11}
\and L.~Mirioni\inst{11}
\and J.~P.~Mittaz\inst{8}
\and C.~Motch \inst{11}
\and J-F.~Olive\inst{4}
\and J.~P.~Osborne\inst{1}
\and C.~G.~Page \inst{1}
\and M.~Pakull\inst{11}
\and B.~H.~Perry \inst{12}$^,$\inst{1}
\and M.~Pierre\inst{2}
\and W.~Pietsch\inst{5}
\and J.~P.~Pye \inst{1}
\and A.~M.~Read \inst{5}
\and T.~P.~Roberts\inst{1}
\and S.~R.~Rosen \inst{8}
\and J-L.~Sauvageot\inst{2}
\and A.~D.~Schwope\inst{7}
\and K.~Sekiguchi\inst{14}
\and G.~C.~Stewart\inst{1}
\and I.~Stewart\inst{1}
\and I.~Valtchanov\inst{2}
\and M.~J.~Ward \inst{1}
\and R.~S.~Warwick\inst{1}
\and R.~G.~West \inst{1}
\and N.~E.~White \inst{15}
\and D.~M.~Worrall\inst{16}
}
 \offprints{M G Watson}
\institute{
	X-ray Astronomy Group, Department of Physics and Astronomy,
        Leicester University, Leicester LE1 7RH, UK
\and
	DSM/DAPNIA/SAp, Bt 709, CEA-Saclay,
        F--91191 Gif-sur-Yvette, France
\and
	Instituto de Fis\'{i}ca de Cantabria (CSIC-UC),
	E--39005 Santander, Spain   
\and
	Centre d'Etude Spatiale des Rayonnements,
        F--31028 Toulouse, France
\and
	Max Planck Institut f\"ur extraterrestrische Physik,
	D-85741 Garching bei M\"unchen, Germany
\and
	Centre for Astrophysics, University of Central Lancashire, 
	Preston, PR1 2HE, UK
\and
	Astrophysikalisches Institut Potsdam (AIP), 
        D--14482 Potsdam, Germany
\and
	Mullard Space Science Laboratory, University College London, Holmbury St.Mary, Dorking RH5
	6NT, UK 
\and
	Harvard-Smithsonian Center for Astrophysics, Cambridge MA 02138, USA
\and
	Institute of Astronomy, University of Cambridge, 
	Cambridge CB3 OHA, UK 
\and
	Observatoire Astronomique de Strasbourg, F--67000 Strasbourg, France
\and
	NASA-GSFC/HEASARC--Raytheon Co.
\and
	Osservatorio Astronomico di Brera,
        I--20121 Milano, Italy
\and
	Subaru Telescope/NAOJ, Hilo, HI 96720, USA
\and
LHEA, NASA-GSFC, Greenbelt, MD 20771, USA
\and
	Dept.of Physics, University of Bristol, Bristol, BS8 1TL,
 UK 
\and
	{\em XMM-Newton} SOC, Villafranca, E--28080 Madrid, Spain\\
}
 \date{Received 2 October 2000; accepted 27 October 2000}
    \titlerunning{The {\em XMM-Newton} Serendipitous Survey}
 
    \authorrunning{Watson et al.}

\abstract{
This paper describes the performance of {\em XMM-Newton} for serendipitous
surveys and summarises the scope and potential of the {\em XMM-Newton}
Serendipitous Survey. The role of the Survey Science Centre (SSC) in the
{\em XMM-Newton} project is outlined. The SSC's follow-up and identification
programme for the {\em XMM-Newton} serendipitous survey is described together
with the presentation of some of the first results. 
}
 
 \maketitle
 
\section{Introduction}
\label{s1}
In contrast with all-sky X-ray surveys which typically provide relatively
shallow coverage, serendipitous surveys provide much deeper observations,
albeit restricted to smaller sky areas. Serendipitous X-ray sky surveys,
taking advantage of the relatively wide field of view afforded by typical
X-ray instrumentation, 
have been pursued with most X-ray astronomy satellites since the {\em Einstein}
Observatory. The resultant serendipitous source catalogues (e.g. EMSS 
[835 sources]
Gioia et al. \cite{gioia}, Stocke et al., \cite{stocke}; WGACAT [$\sim
62000$ unique sources] White, Giommi \& Angelini, \cite{WGA}; {\em ROSAT}
2RXP \& {\em ROSAT} 1RXH [$\sim 95000$ \& $\sim 11000$ sources
respectively]
the {\em ROSAT} Consortium, 2000)
have been the basis for numerous studies and have made a significant
contribution to our knowledge of the X-ray sky and our understanding of
the nature of the various Galactic and extragalactic source populations.

\begin{figure*}[htb!]
\parbox{5.7cm}{\resizebox{\hsize}{!}{\includegraphics{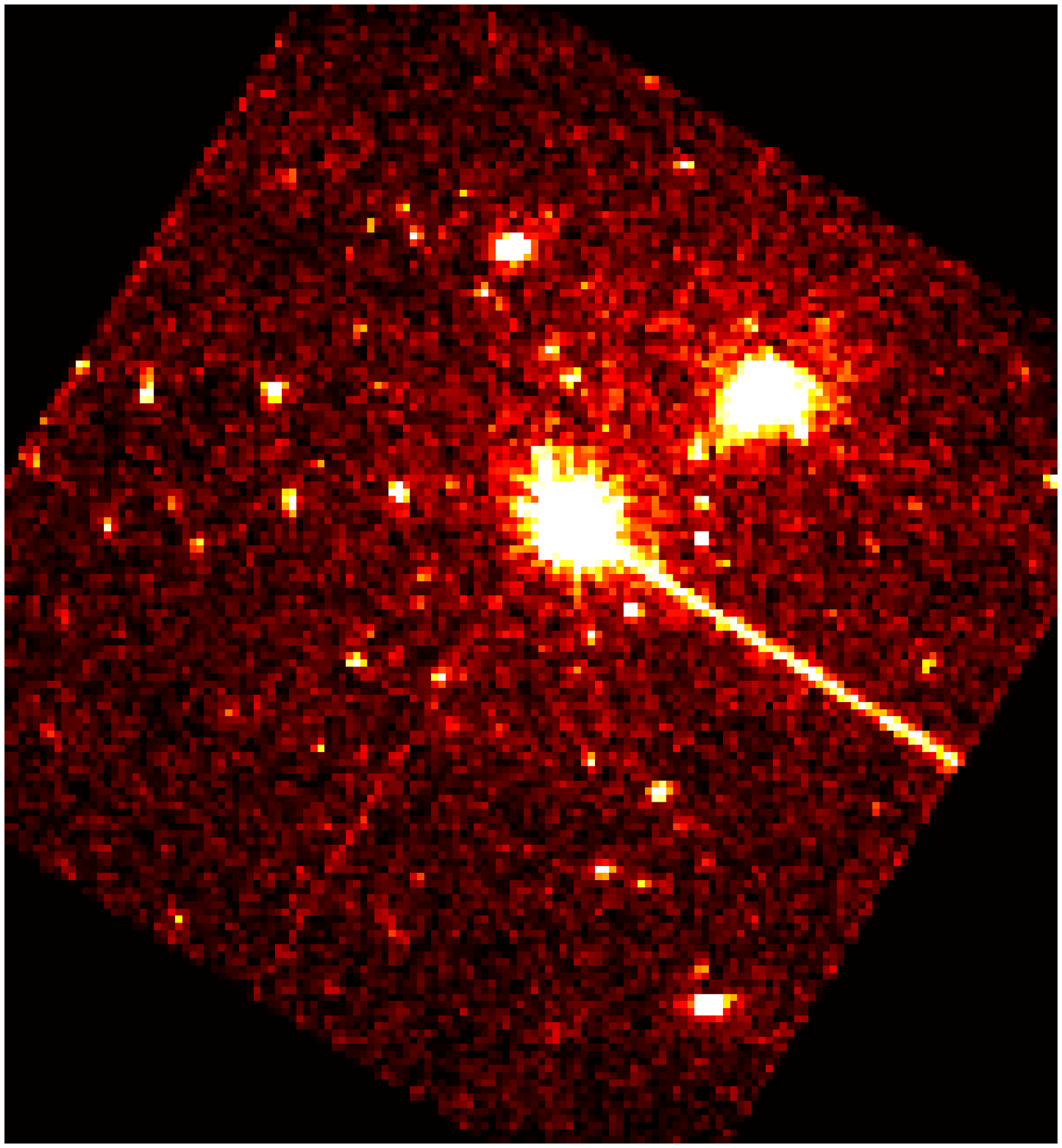}}}
\parbox{6.12cm}{\resizebox{\hsize}{!}{\includegraphics{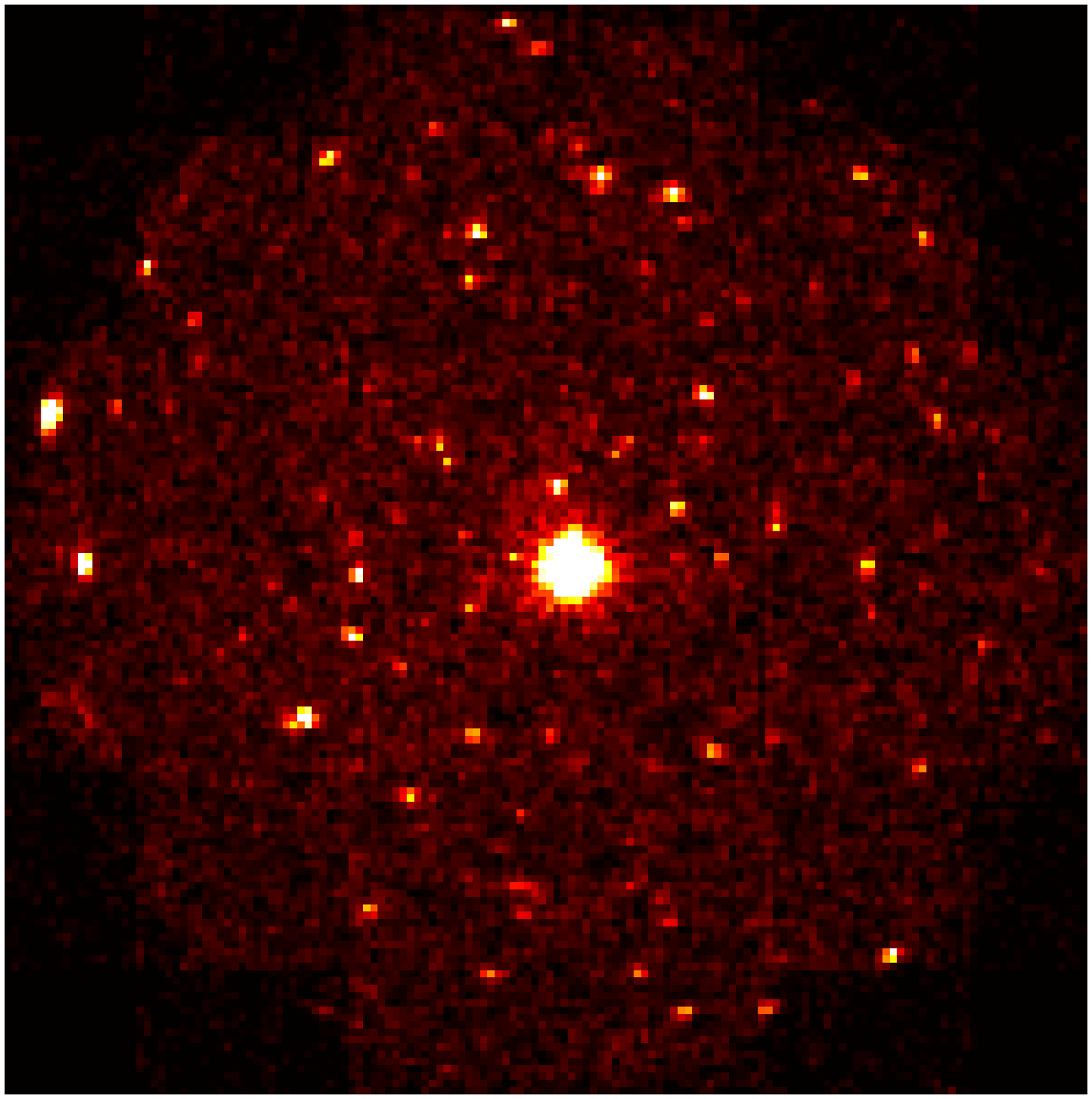}}}
\parbox{5.98cm}{\resizebox{\hsize}{!}{\includegraphics{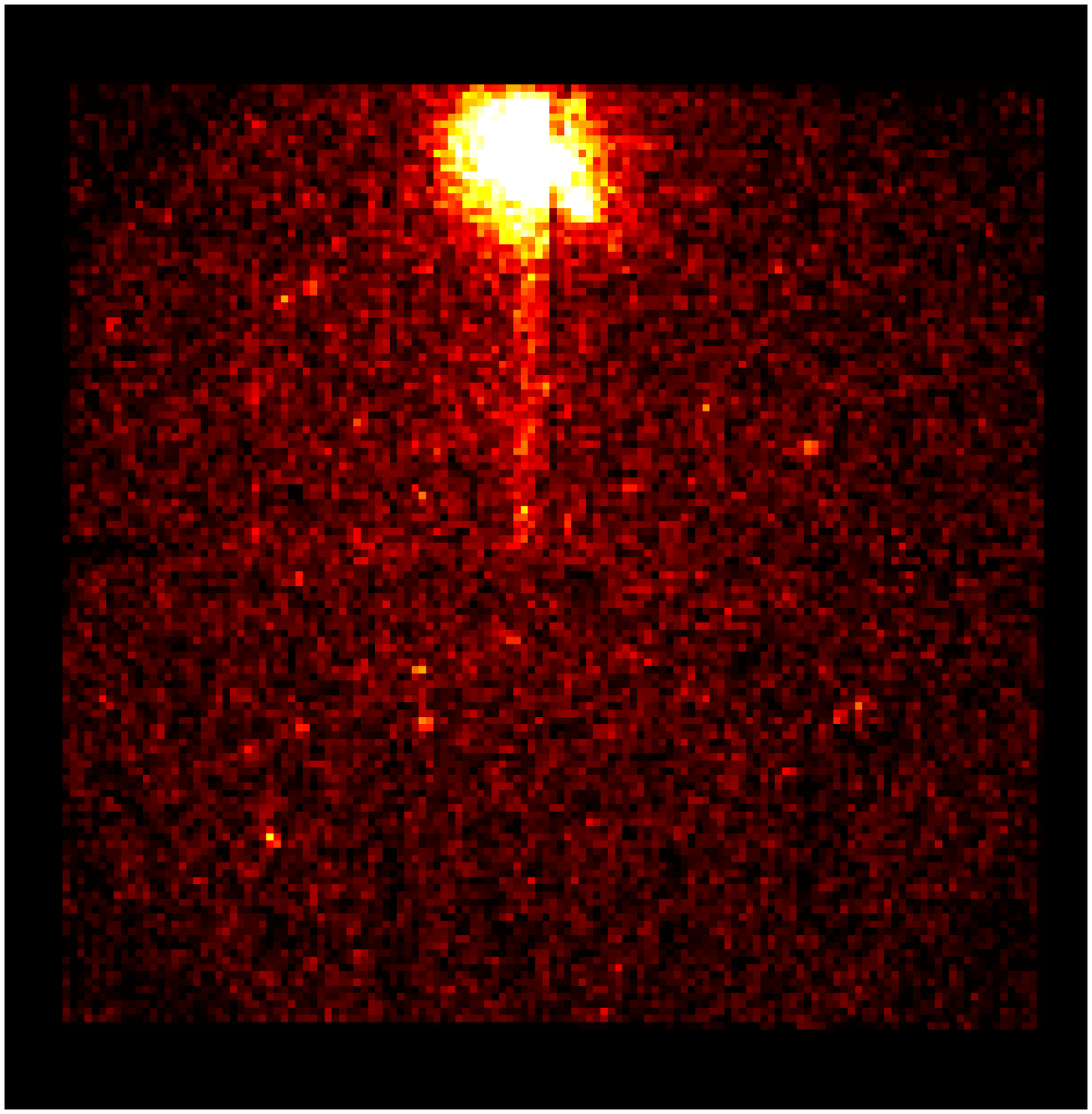}}}
\parbox{5.67cm}{\resizebox{\hsize}{!}{\includegraphics{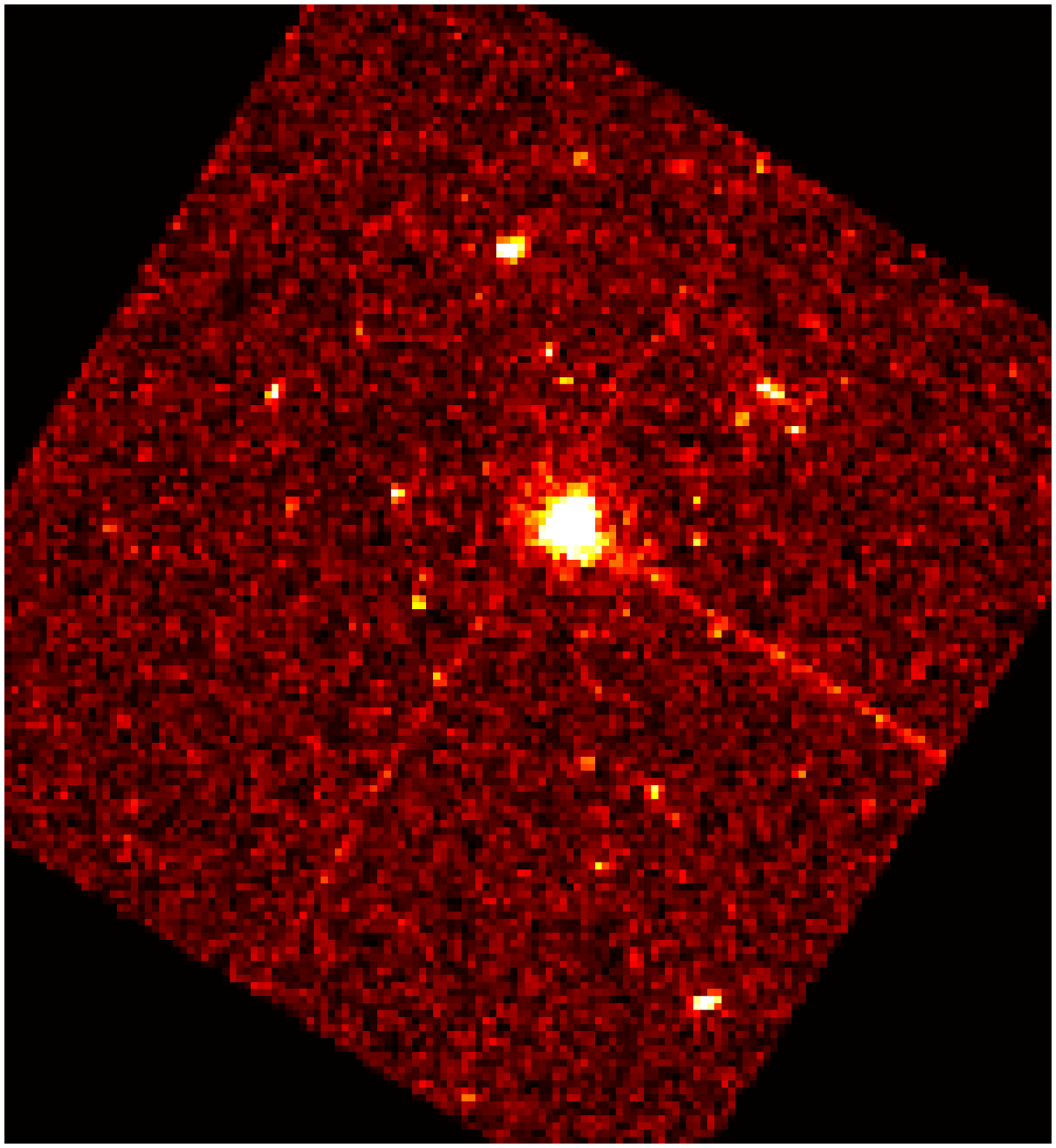}}}
\parbox{6.12cm}{\resizebox{\hsize}{!}{\includegraphics{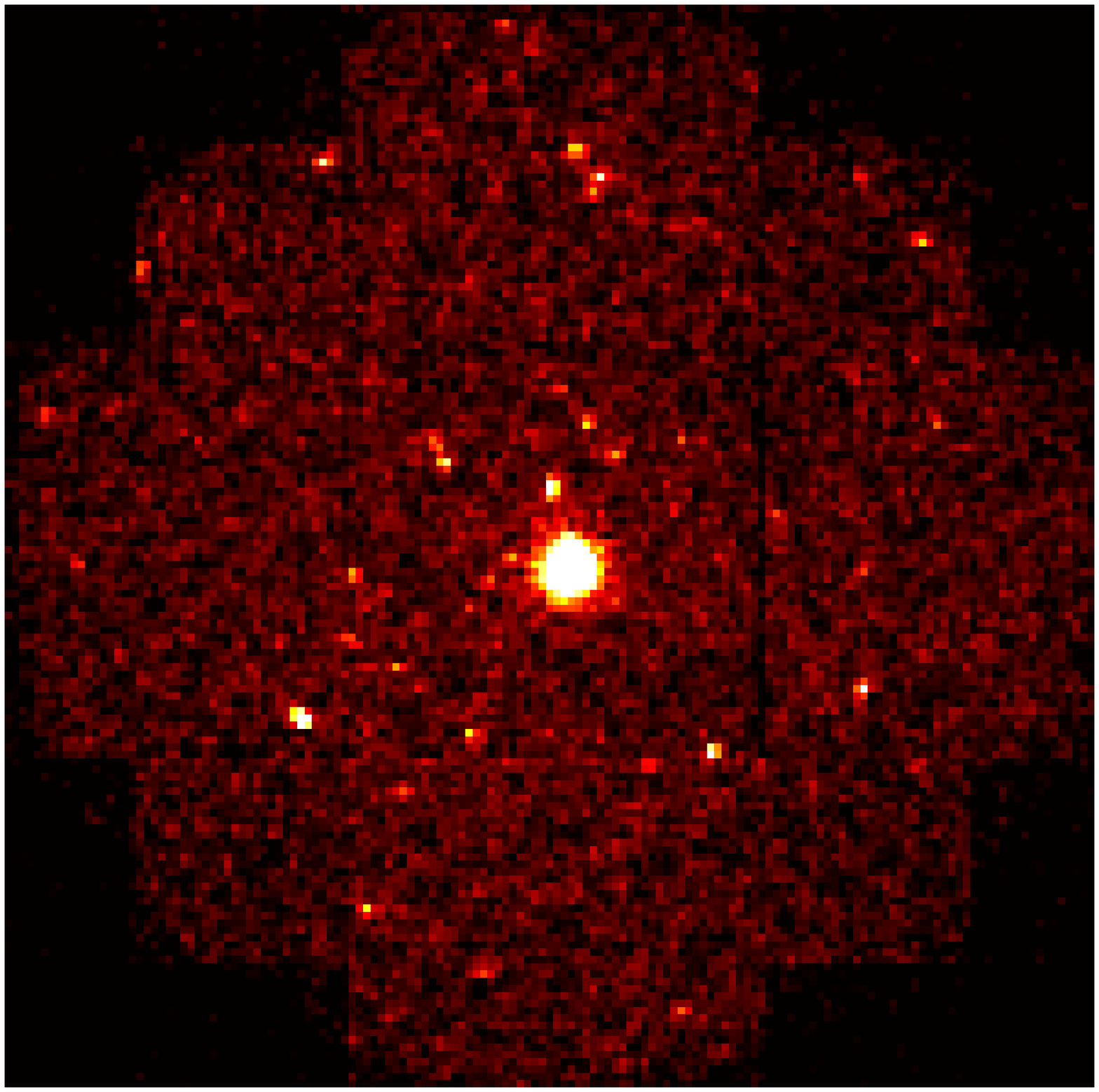}}}
\parbox{5.92cm}{\resizebox{\hsize}{!}{\includegraphics{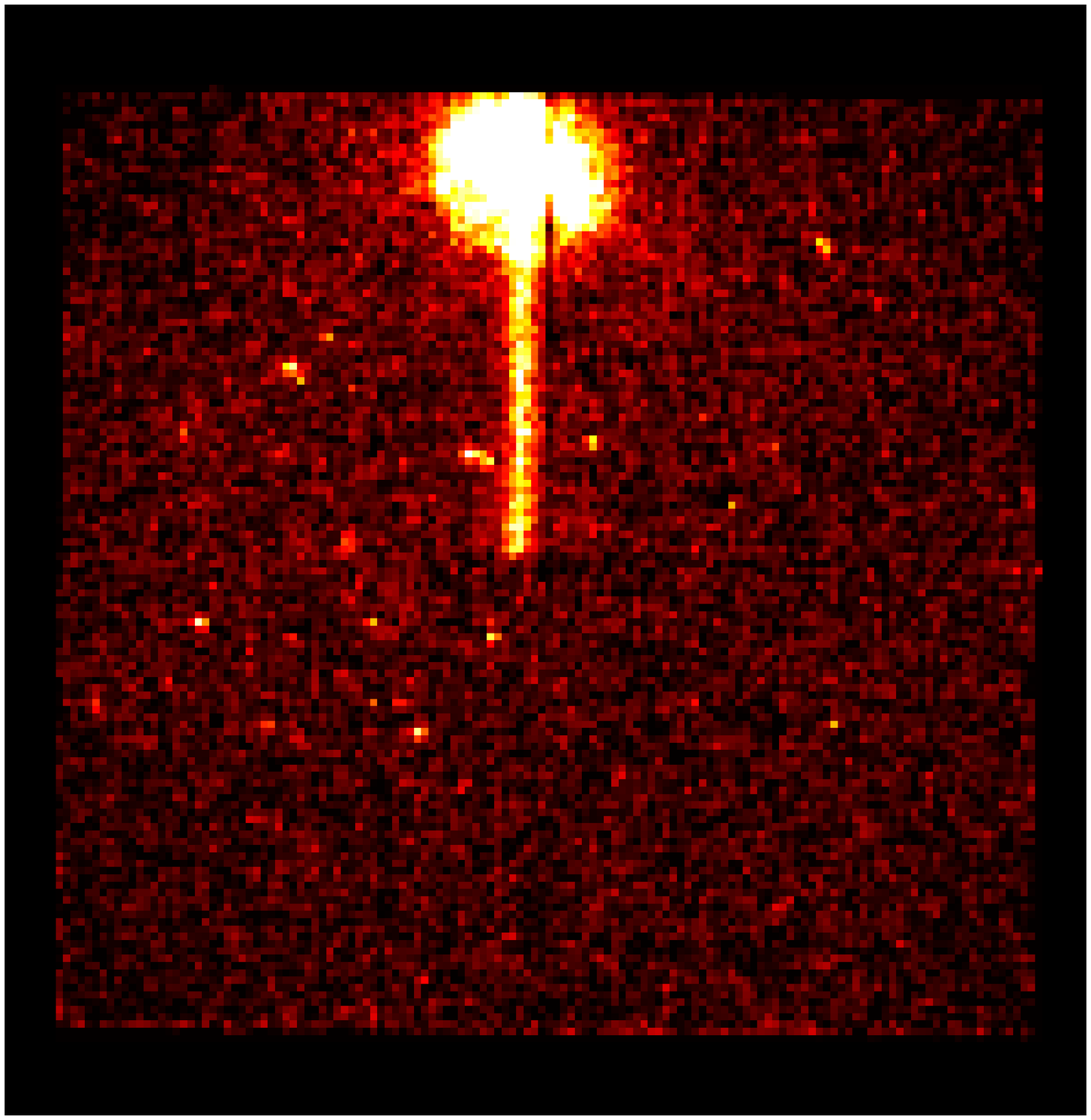}}}%

\caption[]
{EPIC images of three fields. Essentially the whole field, 30 arcmin.
across, is shown in each case. Top row shows the soft band (0.5-2 keV)
images and the bottom row the corresponding hard band (2-10 keV)
images.
{\em
Left:} Mkn 205, EPIC pn image; {\em
Centre:} OY Car, EPIC MOS1+MOS2 image; {\em Right:} G21.5$-$0.9 (offset
pointing), EPIC pn image. Further details are given in Table~\ref{tab1}.
The images have been smoothed to aid the
visibility of faint sources. The trails associated with bright
sources in the pn images are due to `out-of-time' events which arrive
during the read-out cycle of the CCDs and have not been screened out.}
\label{f1}
 \end{figure*}

The {\em XMM-Newton} Observatory (Jansen et al., \cite{jansen}), launched in
December 1999, provides unrivaled capabilities for serendipitous X-ray
surveys by virtue of the large field of view of the X-ray telescopes with
the EPIC X-ray cameras (Turner et al., \cite{turner}; Str\"uder et al.,
\cite{struder}), and the high throughput afforded by the heavily nested
telescope modules. This capability guarantees that each {\em XMM-Newton}
observation provides a significant harvest of serendipitous X-ray sources
in addition to data on the original target. This potential is now
starting to be realised, as is described below.

\section{The role of the {\em XMM-Newton} Survey Science Centre (SSC)}
\label{s2}
In recognition of the importance of the {\em XMM-Newton} serendipitous science
ESA solicited proposals for an {\em {\em XMM-Newton} Survey Scientist} in 1995. The
role of the Survey Scientist centres on providing a coordinated approach
to {\em XMM-Newton} serendipitous data to ensure that the whole scientific
community can exploit this valuable resource. In fact the role of the
Survey Scientist required by ESA was much wider than this, involving the
substantial additional tasks of making a major contribution to the
development of the scientific processing and analysis software for
{\em XMM-Newton}, the routine ``pipeline" processing of all the 
observations and the compilation of the {\em XMM-Newton} Serendipitous
Source Catalogue.

The {\em XMM-Newton} Survey Science Centre (SSC) consortium was selected by ESA
in early 1996. The SSC is an international collaboration involving a
consortium of 8 institutions in the UK, France and Germany, together with
7 Associate Scientists.
The SSC's role in facilitating the exploitation of the {\em XMM-Newton}
serendipitous survey is the main subject of this paper, but here we
emphasise the other contributions that the SSC is making to the project.
Since 1996 the SSC has been working closely with ESA's Science Operations
Centre (SOC) staff in the development of the scientific analysis software
required for the {\em XMM-Newton} project. This is the software that both
carries out the initial pre-processing of the {\em XMM-Newton} scientific data
and permits the detailed scientific analysis of the observations: the
modules can be used in a fixed configuration for the routine processing
of the {\em XMM-Newton} data, and can be used in an interactive configuration
by {\em XMM-Newton} observers to carry out custom analysis of their
data. For documentation of the {\em XMM-Newton} science
analysis software, a collection of software tools now known as the
{\em XMM-Newton} ``SAS", see {\tt http://xmm.vilspa.esa.es/sas/}. At the time
of writing version 5.0 of the SAS is undergoing final testing with a
planned full public release of the system before the end of 2000.

In parallel with the software developments, the SSC has developed the
infrastructure required to carry out, on behalf of ESA, the routine
``pipeline" processing of all the {\em XMM-Newton} observations from each of
the three instruments. The aim here is to provide a set of data products
which will be of immediate value for the {\em XMM-Newton} observer as well as
for the science archive where they will also be stored for
eventual public release. The {\em XMM-Newton} data products include calibrated,
``cleaned" event lists which are intended to provide the starting point
for most interactive analysis of the data as well as a number of
secondary high-level products such as sky images, source lists, 
cross-correlations with archival catalogues, source
spectra and time series. These provide a useful overview of the
observation for the {\em XMM-Newton} observer as well as constituting key
archival resources and the starting point for the compilation of the
Serendipitous Source Catalogue.

The {\em XMM-Newton} Serendipitous Source Catalogue will be based on the EPIC
source lists\footnote{OM results will also be incorporated.} from the
pipe-line processing, but will also contain comprehensive archival
catalogue data, and results from the SSC (and other) follow-up
programme(s) described below. The SSC will emphasize the reliability,
uniformity and usability of the catalogue paying particular attention to the external catalogue correlations and
identification information. The provision of ancillary information such
as sky-coverage and sensitivity estimates is planned to ensure 
the scientific usefulness of the catalogue.

\section{Performance of {\em XMM-Newton} for serendipitous surveys}
\label{s3}
Observations obtained in the early phases of the {\em XMM-Newton} observing
programme have been analysed by the project teams in order to
establish the in-orbit performance of the Observatory and its
instrumentation. Here we concentrate on those aspects of the
{\em XMM-Newton} performance which are important for serendipitous surveys.

\begin{figure}[h!]
\parbox{4cm}{\resizebox{\hsize}{!}{\includegraphics{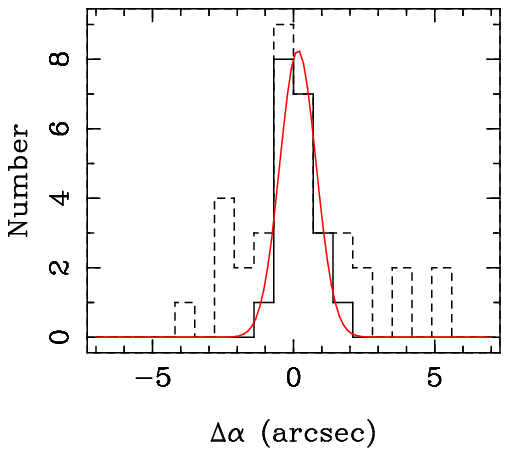}}}
 \hfill
 \parbox{4cm}{\resizebox{\hsize}{!}{\includegraphics{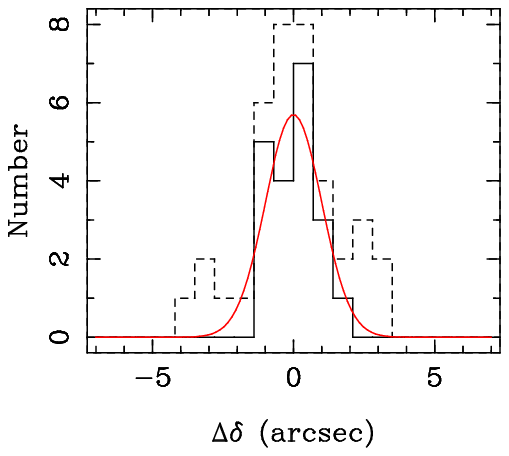}}}
 \caption[]
{Residual position differences between X-ray source positions and
potential optical counterparts extracted from deep i' CCD images obtained
with the INT WFC (see section~\ref{s5}). The solid histogram shows the distribution for
counterparts with a high likelihood of being the correct identification.
The dashed histogram shows the distribution for {\bf all} potential
candidates.  The red curves show the Gaussian fits to the solid histogram
data points.} 
\label{f2} 
\end{figure}

\subsection{Field of view and image quality}
The field of view of the {\em XMM-Newton} X-ray telescopes has a radius of
$\approx 15$ arcmin beyond which baffles reduce the out-of-field response to
very low levels. The X-ray telescope field of view is completely covered
by the EPIC MOS and pn X-ray cameras out to a radius of $\approx 14$ arcmin.
The on-axis point spread functions (PSFs) of the X-ray telescopes,
confirmed by in-orbit data (Jansen et al., \cite{jansen}), have FWHM values 5-7
arcsec and HEW values $\sim$14-15 arcsec. The optical design of the X-ray
telescopes produces a PSF which shows only a very
slow degradation out to a radius of $\sim$10 arcmin (Jansen et al., \cite{jansen}).
With appropriate
on-board and ground processing (as incorporated into the SAS) to remove
hot pixels and noise effects, the EPIC X-ray images are remarkably free
from defects with the most noticeable features being the inter-chip gaps
in the EPIC pn and EPIC MOS cameras. Fig.~\ref{f1} shows example EPIC images
which illustrate these points.

\subsection{Positional accuracy for EPIC source detections}
The accuracy with which EPIC source detections can be measured depends on
the accuracy with which the EPIC camera coordinate system can be related
to the celestial astrometric reference frame and the statistical accuracy
of the individual position determinations in the camera coordinate
system. The transformation between the EPIC camera coordinate system and
sky coordinates provided by the {\em XMM-Newton} aspect system has a typical
accuracy of $\sim$ 4 arcsec. To improve on this, cross-correlation of EPIC
X-ray sources with catalogued optical objects (e.g. the USNO A2.0
catalogue) will be used routinely in the processing pipeline to provide
an updated transformation between the EPIC and sky frames with an expected
accuracy of $\le 1$ arcsec across the whole EPIC field. Using this
approach, the EPIC source positions, even for faint sources close to the
detection limit, have typical 90\% confidence radii of only $\approx 2-5$
arcsec, limited by statistical accuracy of the measurements. An
illustration of the positional accuracy already achieved is given in
Fig.~\ref{f2} which shows the X-ray - optical position residuals for a typical
{\em XMM-Newton} observation. 

\subsection{Survey sensitivity}
The major uncertainty in predicting the EPIC sensitivity pre-launch
was the background levels which would be encountered in-orbit. The
actual in-orbit background levels measured in the first part of the
Lockman Hole observation are $\approx (2.1,2.9)\times 10^{-3}\rm\
ct\rm\ s^{-1}\ {\rm arcmin}^{-2}$ for the EPIC pn camera in the soft
(0.5-2 keV) and hard (2-10 keV) bands respectively. Very similar
values: 
$\approx (2.0,2.6) \times 10^{-3}\rm\ ct\rm\ s^{-1}\ {\rm
arcmin}^{- 2}$ 
are found for the two EPIC MOS cameras combined. A
significant fraction of this background, particularly at hard X-ray
energies, is due to residual particle background with contributions
from both high energy particles interacting with the body of the EPIC
detectors and from soft protons scattered into the focal plane by the
X-ray mirrors. These background levels correspond to the nominal
quiescent background detected outside of background flares at high
galactic latitudes and appear to be relatively stable and reproducible
over the first 9 months of science operations.\footnote{The soft band
(0.5-2 keV) background levels of course show a factor of few change
over the sky due to intrinsic variations in the Galactic foreground,
cf. {\em ROSAT} studies. The hard band background shows much smaller
variation, except at very low Galactic latitudes.} During background
flares the background can increase by large factors ($>10$ is not
uncommon) and portions of these data are of limited use for faint
source detection or the mapping of low surface brightness emission.

\begin{figure}[h!]
\parbox{8.5cm}{\resizebox{\hsize}{!}{\includegraphics{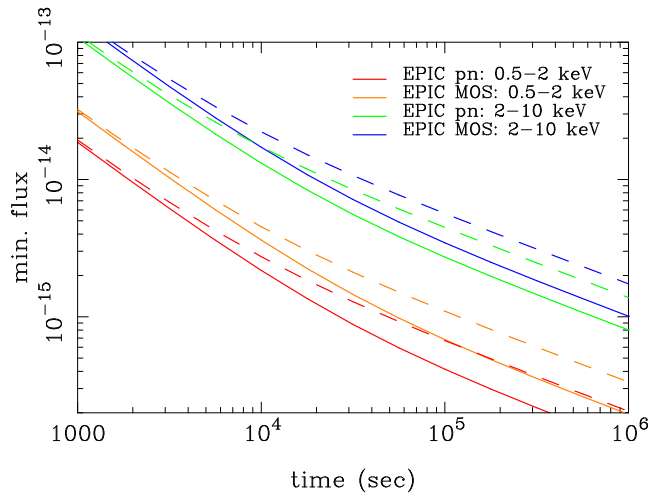}}}
\caption[]
{EPIC sensitivity ($5\sigma$ minimum detectable flux in $\rm erg\rm\
cm^{-2}\rm\ s^{-1}$ in respective bands) as a function of exposure time.
Sensitivity is computed for an assumed $\alpha =1.7$ power-law spectrum
with a column density $N_H = 3\times 10^{20}\rm\ cm^{-2}$. Solid curves are for the nominal background rates quoted in the text.
Dashed curves are for background levels enhanced by a factor 3. The EPIC
MOS curves correspond to the combination of the two cameras.}
\label{f3}
\end{figure}

Using the nominal quiescent background values together with the
measured {\em XMM-Newton} PSF we have computed an updated EPIC point
source sensitivity based on a simple 5$\sigma$ source detection
criterion against assumed purely Poissonian background fluctuations,
as shown in Fig.~\ref{f3}.\footnote{The 5$\sigma$ value represents a
relatively conservative limit which crudely takes into account the
fact that there are additional systematic background effects which
have yet to be characterised in detail. For the effective beam area of
{\em XMM-Newton}, the appropriate limit for purely Poissonian
background fluctuations to yield $\le 1$ spurious source per field is
$\approx 3.5-4\sigma$.} Empirical data from analysis of several {\em
XMM-Newton} fields using the source detection software developed by
the SSC for the SAS are broadly consistent with these plots. The
actual background in an observation depends critically on the fraction
of background flares removed, i.e. the trade-off between net
background levels and net exposure time.  An investigation of a few
example fields demonstrates that the effective sensitivity of typical
observations is within a factor 2 of the values plotted in
Fig.~\ref{f3}.  A few observations are affected by enhanced background
throughout; here the average background can be several times higher
than the nominal values even after the removal of the largest flares.

At very faint fluxes the effective sensitivity is limited by confusion
effects. Although a detailed study of source confusion has not yet
been carried out, the long {\em XMM-Newton} observations of the
Lockman Hole (Hasinger et al., \cite{hasinger01}) demonstrate that
source confusion is not a significant problem in either the soft
(0.5-2 keV) or hard (2-10 keV) X-ray bands for an observation duration
of $\approx 100$ ksec which reaches flux limits $f_X \approx 0.31$ and $\approx 1.4
\times 10^{-15} \rm\ erg\rm\ cm^{-2}\rm\ s^{-1}$ in the soft and hard
bands respectively.

\section{The {\em XMM-Newton} Serendipitous Survey}
\label{s4}
\subsection{Scope and scientific potential}
The high throughput, large field of view and good imaging capabilities
of {\em XMM-Newton} mean that it detects significant numbers of
serendipitous X-ray sources in each pointing, as is illustrated in
Fig.~\ref{f1}. To quantify the serendipitous source numbers expected
from {\em XMM-Newton} observations we note that for the typical
exposure time of 20 ksec, {\em XMM-Newton} will detect sources down to
$\approx 10^{-15}\rm\ erg\ cm^{-2}\ s^{-1}$ in the 0.5-2 keV band, and
$\approx 10^{-14}\rm\ erg\ cm^{-2}\ s^{-1}$ in the 2-10 keV band,
these values referring to a 5$\sigma$ detection in the combined EPIC
pn and MOS cameras (cf. Fig.~\ref{f3}). The X-ray source density at
these flux levels in the high latitude sky is already established from
{\em ROSAT} studies in the soft band (Hasinger et al.,
\cite{hasinger98}) and has now been extended to harder energies by recent
{\em Chandra} and {\em XMM-Newton} studies (Mushotzky et al.
\cite{mushotzky00}; Giacconi et al., \cite{giacconi00}; Hasinger et
al., \cite{hasinger01}). Taking into account the telescope vignetting
which reduces the sensitivity off-axis and the distribution of
exposure times yields an expected $\sim 50-100$ sources per typical
EPIC field (over the whole field, i.e. in a sky area $\approx 0.2$
sq.deg.).  Around 30--40\% of these are expected to be detected in the
hard band, smaller than might have been expected from the higher sky
density at higher energies, this fraction reflecting the lower
effective hard-band sensitivity in flux terms.

\begin{table*}[htb!]
\caption[]{Source content of {\em XMM-Newton} fields}
\begin{tabular}{lcccccc}
Field & \multicolumn{1}{c}{Exposure} & Galactic.&
\multicolumn{4}{c}{Source numbers$^a$} \\
      & \multicolumn{1}{c}{time (ksec)} &latitude      & \multicolumn{2}{c}{soft-band$^b$} &
\multicolumn{2}{c}{hard-band$^b$} \\
  & & & obs & pred & obs & pred\\
\hline
Mkn 205 (EPIC pn data)$^{c}$ & 17          & +42\degr     & 39 & 45 &    22 & 20\\
OY Car (EPIC MOS1+MOS2 data) &    50       &   $-12$\degr        &  64 & 55 &  25 & 30      \\
G21.5$-$0.9 (EPIC pn data)$^{d}$ &   30        & $-1$\degr     &  15 & (0)$^e$ & 19 &
($\sim10$)$^e$     \\
\end{tabular}\\[4mm]
a: Approximate observed and predicted source numbers ($>5\sigma$) for off-axis
angle $\le 12$ arcmin. \\
b: Soft-band: 0.5-2 keV; hard-band: 2-10 keV.\\
c: First exposure in orbit 75 only.\\
d: Offset pointing (cf. Fig.~\ref{f1}).\\
e: Background extragalactic sources only, note that source 
count predictions depend sensitively\\
on the Galactic
column density for this field.\\
\label{tab1}
\end{table*}

The extent to which these expectations are matched by the real data is
illustrated by the images in Fig.~\ref{f1} which show three typical {\em
XMM-Newton} EPIC observations of calibration and performance
verification fields covering a range of Galactic latitudes and
exposure times. The source content of these fields is summarised in
Table~\ref{tab1} (note that the numbers are given for off-axis angles
$\le 12$ arcmin., i.e. an area of $\approx 0.125$ sq.deg.; the full
field numbers are 25--30\% higher). The source numbers in the high
latitude and mid-latitude fields are in general accord with our
expectations (predicted numbers in Table~\ref{tab1} are for the
Hasinger et al., \cite{hasinger98} \& Giacconi et al.,
\cite{giacconi00} source counts and the sensitivities given in
Fig.~\ref{f3} with an approximate correction for the sensitivity
decrease off-axis due to telescope vignetting; these predicted numbers
are only accurate to 10--20\%). The G21.5$-$0.9 field, 
at low Galactic latitude and with a high column density ($N_{HI} +
N_{HII} \approx 
10^{23}\rm\ cm^{-2}$), samples a rather different source
population as shown by the somewhat lower source density and the fact
that there is little correspondence between the soft and hard band
detections. In the soft band the detections in this field are likely
to be dominated by foreground objects, most of which will be active
stars (see section~\ref{s5}). In the hard band the faint source
population revealed is likely to be dominated by a mixture of
background extragalactic sources seen through the Galactic disk and
distant (and hence similarly absorbed) luminous Galactic objects which
may include cataclysmic variables and other low luminosity accreting
systems. The approximate calculation of the expected extragalactic
source numbers in the G21.5$-$0.9 field given in Table~\ref{tab1}
(sensitive to the column density distribution) indicates that a
significant fraction, but not all, of the hard band sources maybe background
objects.

With the current operational efficiency, {\em XMM-Newton} makes of the order
500-800 observations per year, covering $\sim 100$ sq.deg. of the sky.
The {\em {\em XMM-Newton} serendipitous X-ray catalogue} will thus grow at a
rate of $\sim 50000$ sources per year, i.e. the annual rate will be
comparable in size to the complete {\em ROSAT} All Sky Survey, but will go to
fluxes 2-3 orders of magnitude fainter. The catalogue will thus
constitute a deep, large area sky survey which will represent a major
resource for a wide range of programmes. The extended energy range of
{\em XMM-Newton}, compared with previous imaging X-ray missions such as {\em ROSAT}
and the {\em Einstein} Observatory, will mean that {\em XMM-Newton} is expected to
detect significant numbers of obscured and hard-spectrum objects (e.g.
obscured AGNs, heavily absorbed Galactic binaries) which are absent from
earlier studies. This will provide an important extra dimension to the
serendipitous catalogue.

{\em Chandra} observations will also provide a serendipitous sky survey
comparable in many ways to what {\em XMM-Newton} can offer. 
Nevertheless a number of factors, in particular the smaller
field of view with good imaging quality and the fact that the full ACIS
array is not always selected by the observer, 
contribute to producing a significantly smaller sky coverage (by a factor
5-10) for the {\em Chandra} serendipitous sky survey. {\em XMM-Newton} also has
a very significant advantage at photon energies above 4-5 keV: at low
energies the ratio of EPIC/ACIS effective areas is 3-4 but this increases
to $\sim 6$ at 5 keV and $>10$ at 7 keV. The {\em XMM-Newton} serendipitous sky
survey is thus likely to contain much larger samples of highly obscured
sources, an important factor for the study of the objects currently
believed to make up the bulk of the X-ray background (cf. Lockman Hole
study of the 5-10 keV source counts, Hasinger et al., \cite{hasinger01}).

X-ray selection of samples provides a well-proven and extremely efficient
means of finding some of the most astrophysically interesting objects,
many of which have their peak luminosity in the X-ray band. {\em XMM-Newton},
with its broad-band and high sensitivity will provide less biased samples
than those based on previous soft X-ray missions. It is clear that the
{\em XMM-Newton} serendipitous source catalogue will prove to be a major
resource in many areas of scientific investigation. Examples include:

\begin{itemize}
\item the importance of obscuration in the faint AGN population
 \item the evolution of the
 quasar luminosity function  with redshift 
 
 \item the nature
 of faint X-ray galaxies - their contribution to the X-ray
 background
 
 \item the evolution and luminosity function of clusters of galaxies
 and the nature of density fluctuations in the early Universe
 
 \item the coronal
 activity in stars and its dependence on luminosity, spectral type,
 stellar rotation and age
 
 \item the space density of accreting binary systems. 
 \end{itemize}

\subsection{Need for follow-up studies}
In order to exploit the full potential of the {\em XMM-Newton} serendipitous
survey in the context of a wide range of scientific programmes, the key
initial step will be the `identification' of the X-ray sources, i.e. a
knowledge of the likely classification into different object types. For
the {\em XMM-Newton} serendipitous sources, the X-ray observations themselves
will provide the basic parameters of each object: the celestial position,
X-ray flux for all sources and information on the X-ray spectrum, spatial
extent and temporal variability for the brighter objects detected. This
information alone may, in some cases, be sufficient to provide a clear
indication of the type of object, but for the vast majority of sources,
additional information will be required before a confident classification
of the object can be made. Some of this information will come from
existing astronomical catalogues, or from existing or planned large-scale
optical, IR and radio surveys (e.g. SDSS, 2MASS, Denis, FIRST/NVSS), but
it is clear that the full exploitation of the {\em XMM-Newton} serendipitous
survey will require a substantial programme of new observations,
primarily in the optical and IR using ground-based facilities.

\subsection{The SSC Follow-up (XID) Programme}
The overall aim of the SSC Follow-up Programme, for simplicity the `XID
Programme', is to ensure that the potential of the {\em XMM-Newton}
serendipitous survey can be fully exploited by the astronomical
community. The XID Programme is thus designed to maximise its value for a
wide range of potential scientific uses of the serendipitous data. In
order to ensure the optimum utility of the programme, the results from
the SSC XID Programme will enter the public {\em XMM-Newton} Science Archive.\footnote{The SSC XID Programme is restricted to {\em XMM-Newton} data which
is in the public domain, i.e. outside the 1-year proprietary period,
except for observations where the {\em XMM-Newton} observer has given explicit
permission for the SSC to commence follow-up work in the proposal for
{\em XMM-Newton} observations.} The value of building a `statistically'
identified catalogue of EPIC sources and delivering it to the community
is that it will provide large homogeneous samples
for studying class properties and to search for rare objects.

The approach planned by the SSC is a programme which brings together the
{\em XMM-Newton} data themselves, existing catalogue and archival material and
new ground-based observational data in an integrated fashion. The main
new observational elements, the `Core Programme' and the `Imaging
Programme', are outlined below.

\subsection{The XID Core Programme}
The aim of the Core Programme is to obtain the identifications for a
well-defined sample of X-ray sources drawn from selected {\em XMM-Newton}
fields, primarily using optical/IR imaging and spectroscopy. Imaging is
required both to locate potential candidates accurately and reveal their
morphology, whilst the optical spectroscopy provides the diagnostics
needed both for object classification and for determining basic object
parameters such as redshift and spectral slope. The principal objective
is to obtain a completely identified sample which can be used to
characterise the {\em XMM-Newton} source population overall sufficiently well that
we can use the basic X-ray and optical parameters to assign a
`statistical' identification for a large fraction of {\bf all} the
sources in the {\em XMM-Newton} serendipitous source catalogue.

\begin{table*}[htb!] 
\caption[]{XID Core Programme sample parameters} 
\begin{tabular}{llrrc} 
Sample & Flux range$^a$ & Sky dens.$^b$ & \# EPIC$^c$ & R mag.$^d$\\
\hline 
{\small FAINT} &$\ge 10^{-15}$ & $2200$ & 5-10 & 23-25\\ 
{\small MEDIUM} &$\ge 10^{-14}$ & 340& 30-50 & 21-23\\ 
{\small BRIGHT} &$\ge 10^{-13}$ & 10 & 1000 & 17- 21\\ 
{\small GALACTIC}& $\geq$ 5 10$^{-15}$ & $\sim$ 300  & 40 & wide\\
        &                     &             &    &range\\
\end{tabular} 

a: X-ray flux in $\rm erg\rm\ cm^{-2}\rm\ s^{-1}$ in the 0.5-4.5 keV band\\
b: source density (deg$^{-2}$)\\
c: no.of EPIC fields required\\
d: expected R-magnitude range of counterparts\\
\label{tab2}
\end{table*} 

The strategy involves two samples: one for the high and one for the low
galactic latitude sky. The high galactic latitude sample consists of
three subsamples, each containing $\approx 1000$ X-ray sources in three
broad flux ranges: $f_x > 10^{-15}$ ; $f_x > 10^{-14}$; $f_x >
10^{-13}\rm\ erg\ cm^{-2}\ s^{-1}$ in the 0.5-4.5 keV band (see
Table~\ref{tab2}).
The size of the subsamples is dictated by the need to identify enough
objects to reveal minority populations. Studying sources at a range of
X-ray fluxes is necessary because we already know that the importance of
different source populations changes with X-ray flux level. A parallel
study is planned for the low galactic latitude sample. 
Taking into
account the difficulty of obtaining completely identified samples close
to the Galactic plane, here we aim to identify a sample of $\sim 1000$
sources above a flux level of $\approx 5\times 10^{-15}\rm\ erg\ cm^{-2}\
s^{-1}$(0.5-4.5 keV), from target fields covering a range of galactic
latitudes and longitudes (Table~\ref{tab2}).
In our
programme we also need to minimise bias against any particular X-ray
spectral form by recognising flux limits in different X-ray bands. To
achieve this, additional soft and hard X-ray selected samples will be
included in the programme.

The requirements for the Core Programme are summarised in Table~\ref{tab2}.  The
median optical magnitudes in Table~\ref{tab2} (and the size of the subsamples)
illustrate that this is a very demanding programme. For example
spectroscopy of the counterparts to the faintest X-ray sources will
require access to multi-object spectrographs on 8-10 m class telescopes,
and indeed a substantial fraction of counterparts will be inaccessible,
whilst even for the sources in the `medium' flux sample access to 4-8 m
class telescopes will be needed. For the `bright' sample the situation
will be somewhat different in that a large fraction of counterparts will
be on existing optical sky survey material (e.g. POSS), and a significant
fraction may have likely identifications with catalogued objects, leading
to reduced requirements for spectroscopic follow-up.  

Taking into account the fact that reaching the faint sample depth
requires long {\em XMM-Newton} exposures, and that most of these observations
are being pursued for surveys by the observation PIs, the current
emphasis in the SSC XID programme is on the medium flux sample (and later
as the observations become available, on the bright sample) which,
although ambitious, is feasible provided access to the necessary
ground-based observing facilities is possible.

\subsection{The Imaging Programme}
Whilst the Core Programme focuses on the identification of a selected
subsample of {\em XMM-Newton} serendipitous sources, the complementary Imaging
Programme aims to obtain optical/IR photometry and colours for a large
number of {\em XMM-Newton} fields. The programme rationale is based on the fact
that a combination of X-ray flux \& X-ray colours (from the {\em XMM-Newton}
data) and optical magnitude and optical colours (e.g. from new
ground-based observations) will provide the key parameters which make
possible an accurate `statistical' identification of the {\em XMM-Newton}
sources. This will be possible using the results from the Core Programme
which characterise the {\em XMM-Newton} source populations, thus providing the
link, in a statistical sense, between the source identification and these
basic parameters. 

Multi-colour optical imaging provides good discrimination between object
types (e.g. AGN-star separation) as well as photometric redshifts, whilst
IR imaging has an important role to play as the counterparts to obscured
X-ray sources are expected to also show significant reddening.

The Imaging Programme will primarily be pursued using new optical/IR
imaging. To reach R$\sim 23-25^{\rm m}$, and K$\sim 20^{\rm m}$, the typical
values required for the serendipitous survey follow-up, is within the
range of 2-4m class telescopes.\footnote{Imaging from the {\em XMM-Newton} OM
(Mason et al., \cite{mason}) will in many cases provide valuable data for
the Imaging Programme, in particular by extending the coverage to the UV.
Availability of appropriate OM data cannot be guaranteed however, because
the choice of OM readout modes, filters etc. are selected by the
observation PI.}

\section{XID Programme implementation and first results}
\label{s5}
The SSC XID Programme started in April 2000 with initial optical and
IR imaging and the first spectroscopic data were taken in June 2000.
As noted above, the current emphasis of the programme is on the medium
flux sample (cf. Table~\ref{tab2}). A substantial fraction of the
observing time for the current programme was awarded within the Canary
Islands International Time Programme (ITP) to a project, known as
``AXIS" ({\em An {\em XMM-Newton} International Survey}, {\tt
http://www.ifca.unican.es/$\sim$xray/AXIS}) which focuses on the
follow-up of the medium flux XID sample and the nature of the hard
X-ray source population.

\begin{figure*}[hbt!]  
\parbox{5cm}{\resizebox{\hsize}{!}{\includegraphics{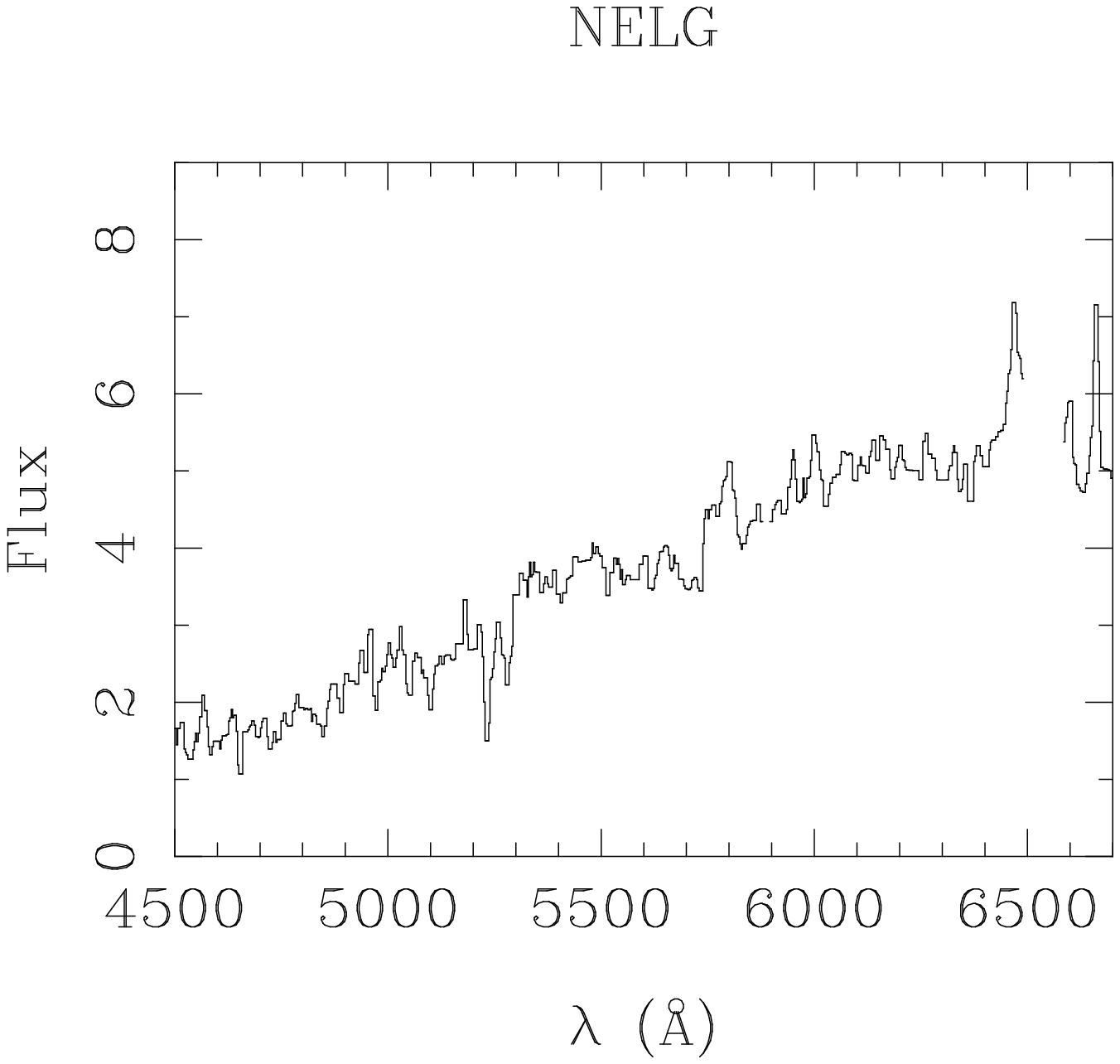}}} 
\hfill
\parbox{5cm}{\resizebox{\hsize}{!}{\includegraphics{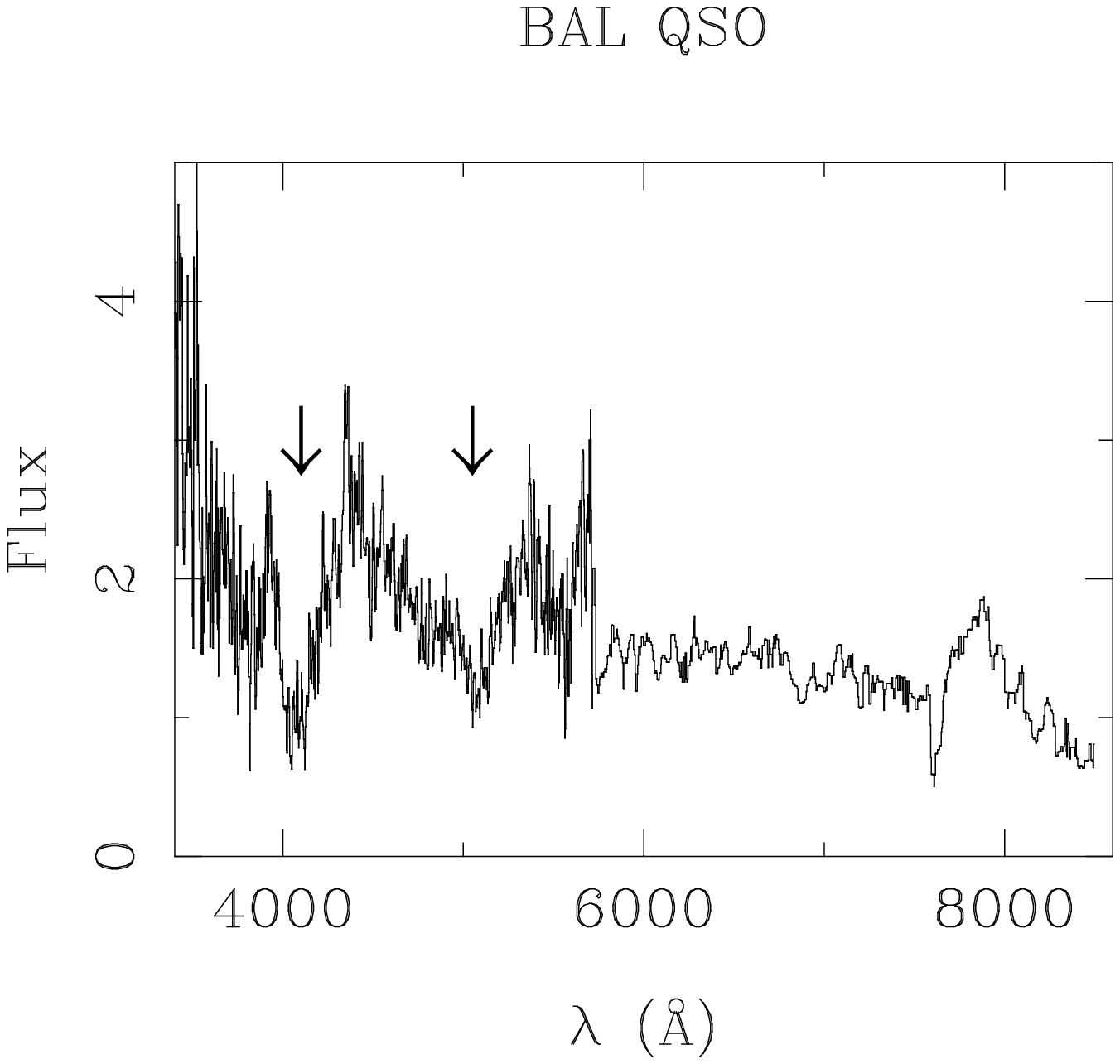}}} 
\hfill
\parbox{5cm}{\resizebox{\hsize}{!}{\includegraphics{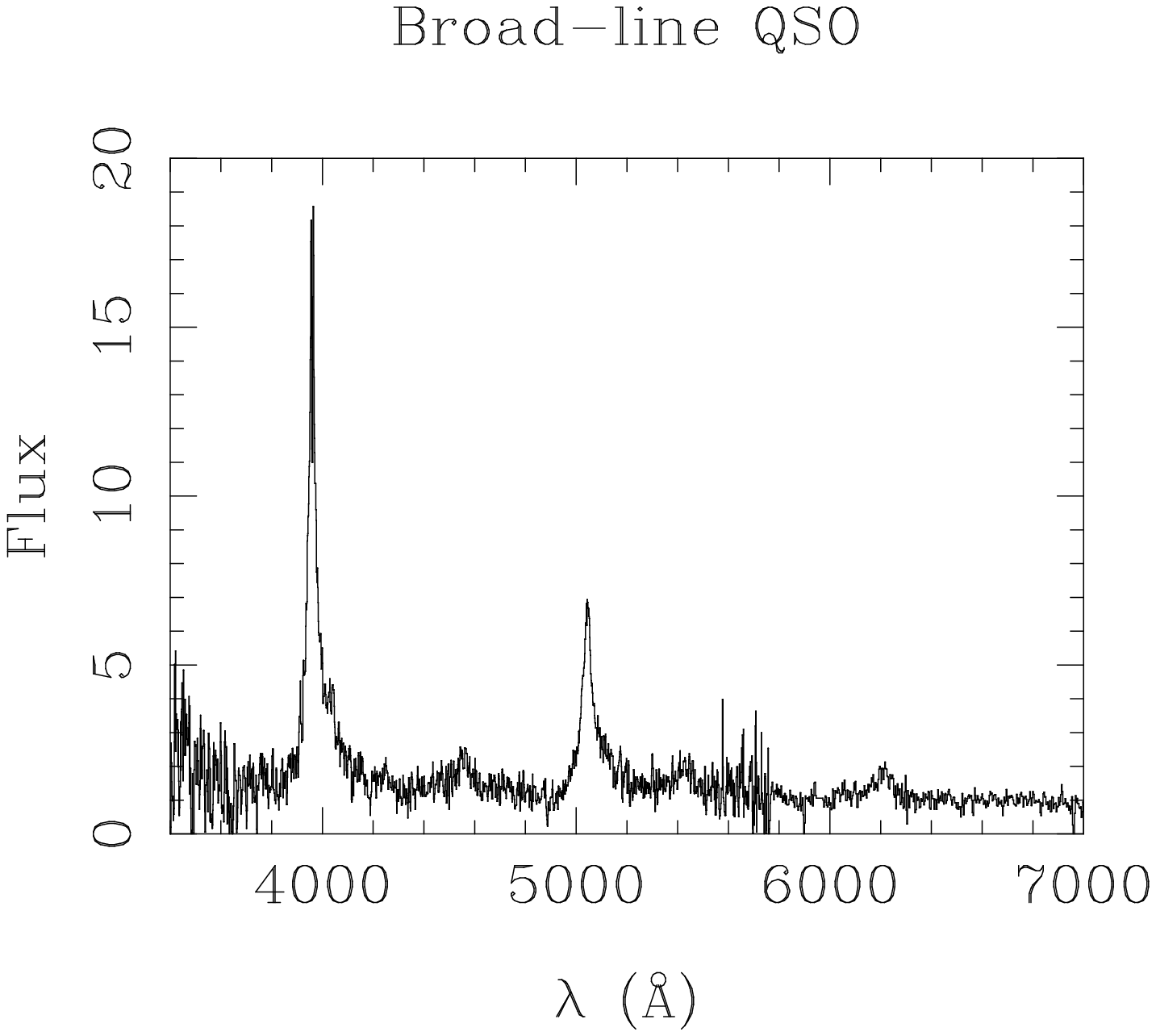}}}\\[4mm]
\hspace*{6mm}
\parbox{4cm}{\resizebox{\hsize}{!}{\includegraphics{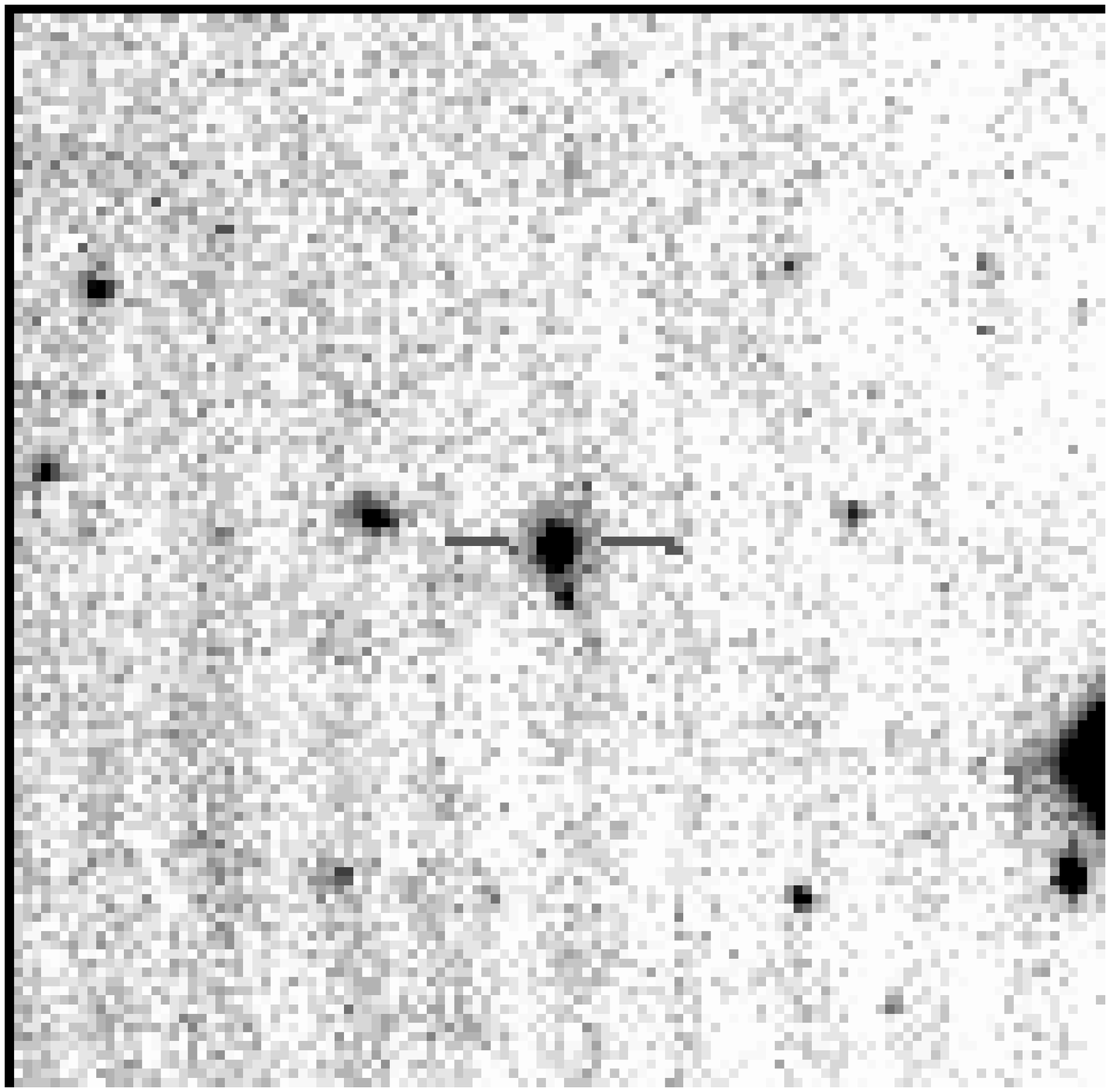}}} 
\hfill
\parbox{4cm}{\resizebox{\hsize}{!}{\includegraphics{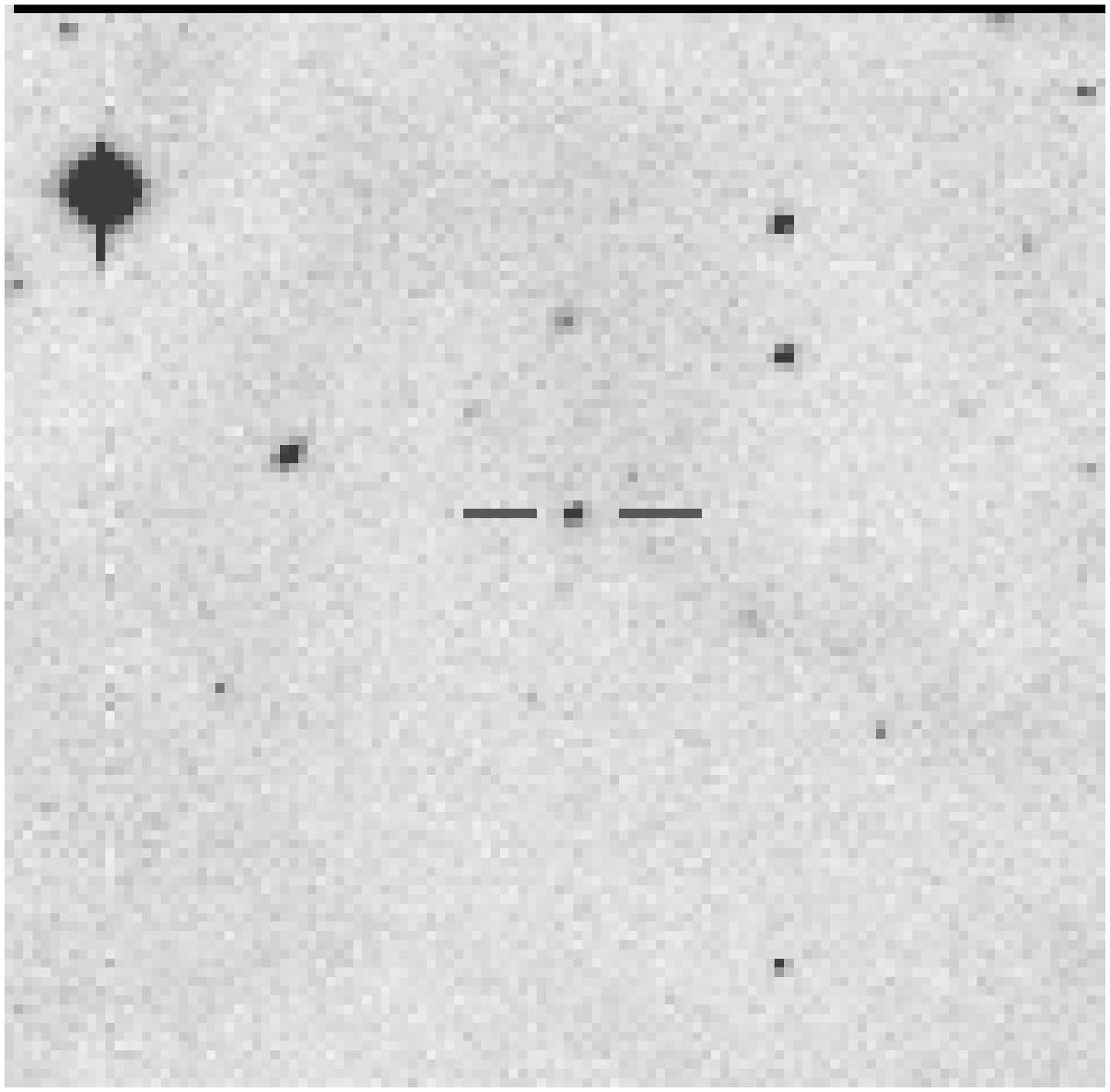}}} 
\hfill
\parbox{4cm}{\resizebox{\hsize}{!}{\includegraphics{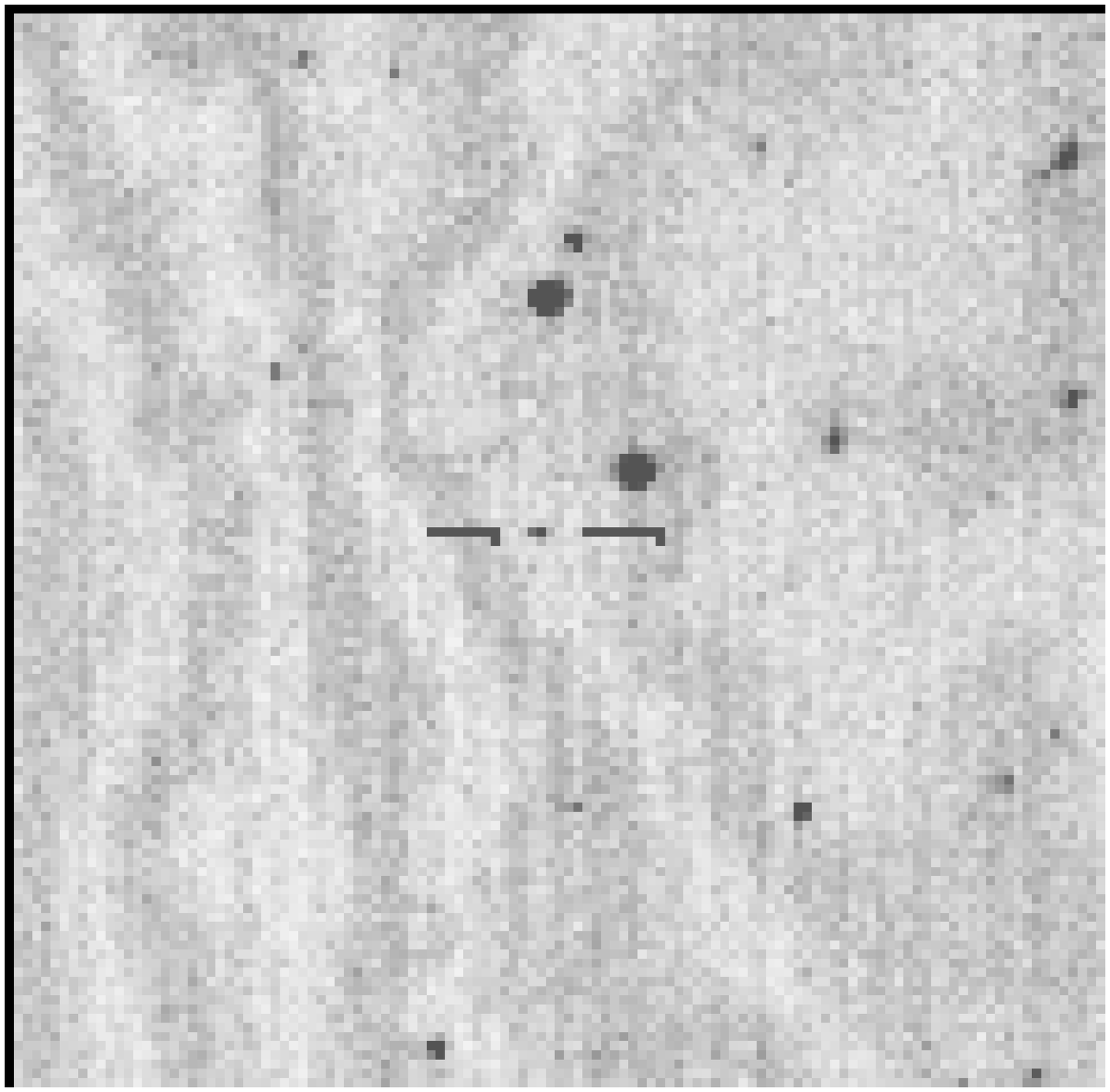}}}%
\hspace*{3mm}
\caption[]
{Optical spectra and finding charts of the counterparts for three 
serendipitous {\em XMM-Newton} sources. {\em Left panels:} an
object in the Mkn 205 field identified with a $z = 0.33$ galaxy.
{\em Centre panels:} an object in a GT field
identified with a BAL quasar at $z=1.82$. {\em Right panels:} an object in
the same field
identified with a quasar at $z=2.26$.   
The NELG spectrum is from the WHT + WYFFOS (gaps due to poor
sky subtraction) and the lower two from WHT + ISIS (both arms
shown). Optical fluxes are in units of  
$10^{-17}\rm\ erg\rm\ cm^{-2}\rm\ s^{-1}$\AA$^{-1}$. Finding
charts are $\approx 2\times 2$ arcmin. in size and are taken from
INT WFC i'-band imaging.}
\label{f4}
\end{figure*}

Optical imaging has been obtained for more than 30 {\em XMM-Newton}
fields using the INT Wide Field Camera and the ESO-MPG 2.2m Wide Field
Imager. The Sloan u,g,r,i,z bands have been chosen for optical imaging
as these provide significant advantages for the determination of
spectroscopic redshifts. Near-IR imaging of a few fields has already
been obtained with the INT 2.5m and the CIRSI IR camera in the H band
(CIRSI does not offer the K band).  For spectroscopy, three short runs
have already been completed using the WHT 4.2m with the WYFFOS
multi-object spectrograph, the WHT with the ISIS dual-beam
spectrograph and with the NOT 2.5m with the ALFOSC spectrograph. The
WHT spectroscopy with WYFFOS can reach R$\sim 21^{\rm m}$ and $\sim
22^{\rm m}$ with ISIS. The NOT/ALFOSC spectroscopy concentrated on
objects with R$<18^{\rm m}$. 

We have already obtained spectroscopic identifications for $\sim 40$
{\em XMM-Newton} sources in the first few fields studied. These fields
include the Mkn 205 field (cf. Fig.~\ref{f1} and Table~\ref{tab1}, we
used detections from the combined data of the two exposures), three of
the five G21.5$-$0.9 calibration observations (these cover $\sim 0.5$
sq.deg; cf. Fig.~\ref{f1} and Table~\ref{tab1} which relate to one of
the five observations), and two Guaranteed Time (GT) fields at high
galactic latitude. Fig.~\ref{f4} shows the finding charts and optical
spectra of three example identifications made in these fields:
\begin{itemize}
\item A $z=0.33$ galaxy in the Mkn 205 field showing H$\beta$ and [OIII] 
narrow emission lines at $\lambda\lambda
6466, 5803, 6595$\AA. Judging by its X-ray luminosity ($L_X
\approx 10^{43}\rm\ erg\rm\ s^{-1}$) and line ratios this galaxy probably
hosts an active nucleus.
\item
An object at $z=1.82$ in a GT field\footnote{A field
for which SSC follow-up permission was granted by the observation PI.} which shows broad emission
lines (notably Si IV, C IV, CIII] \& Mg II at $\lambda\lambda 3939,
4365, 5381, 7890$\AA) with broad blue-shifted absorption troughs to
the CIV \& CIII] lines (arrowed in Fig.~\ref{f4}). This thus appears
to be a broad absorption line (BAL) quasar. If confirmed this would be
one of the first X-ray selected BAL objects so far, such objects being
extremely faint at soft X-ray energies (Green \& Mathur,
\cite{green}).
\item An object at $z=2.26$, in the same field, showing the
typical broad emission lines of a quasar. The strongest lines are
Ly$\alpha$, CIV \& CIII] at  $\lambda\lambda 3959, 5040, 6212$\AA.
\end{itemize}

Although it is premature to discuss the results in any detail at this
stage, as our programme has yet to accumulate statistically useful
samples of identifications, we can nevertheless make some general
remarks. In the high galactic latitude fields for which we have
spectroscopic data so far, broad-line AGN (15 objects) dominate the
total number of identified sources. The other identified objects
include 4 narrow emission line galaxies (NELGs; in this context this
means extragalactic objects with no obviously broad lines), 1
``normal" galaxy and 3 stars (2 of which are dMe).

As found in previous {\em ROSAT} surveys, e.g. Lehmann et al.,
(\cite{lehmann}), all the NELGs are at $z<0.5$. Some of these NELGs
may actually be obscured AGN, but more thorough analysis and better
quality spectra are required to confirm this. Our programme has not
yet found an example of a high redshift type II AGN, but this is as
expected because we have concentrated on the X-ray and optically
brighter objects so far.

In the one low latitude region for which we have spectroscopy so far
(the three overlapping G21.5$-$0.9 fields), initial problems in
establishing good astrometry for the field led to poorer X-ray source
positions. Because of the high star density at low latitudes, this
exacerbated the difficulty of selecting optical counterparts.
Accordingly our observations to date have concentrated on the brighter
optical counterparts. Of the 27 sources observed to date, 11 show
clear stellar spectra. Only 3 of these show clear spectroscopic
evidence of being active at the current spectroscopic resolution; this
highlights the fact that {\em XMM-Newton} can detect stars with lower
activity. Analysis of the data for the other objects observed is still
in progress. Given the current emphasis on the brighter counterparts,
it is not surprising that our study has not yet revealed other classes
of object, e.g. distant, faint cataclysmic variables or even
background AGN which will be both faint and highly reddened.

The XID programme is in the early stages of what is a long-term
project.  We look forward to realising the potential of the {\em
XMM-Newton} serendipitous survey over the coming years in what seems
likely to be a rich and rewarding programme.

\begin{acknowledgements}
It is a pleasure to acknowledge the efforts and dedication of the
large team of people who have been essential in developing the SSC,
not all of whom are represented as co-authors of this paper. We
gratefully acknowledge allocations of telescope time for the public
XID programme by ESO, PATT and the CCI for the AXIS project. Mat Page
and Ismael Perez-Fournon are thanked for their contribution to the
AXIS programme observing. The INT and the WHT are operated on the
island of La Palma by the Isaac Newton Group in the Spanish
Observatorio del Roque de Los Muchachos of the Instituto de
Astrofis\'{i}ca de Canarias.

Support for the SSC project provided by the following agencies:
Deutsches Zentrum f\"{u}r Luft- und Raumfahrt (DLR) under grant
numbers 50 OR 9908 0 \& 50 OX 9701 5; PPARC (UK); CNES (France) and
NASA (via the HEASARC). TM thanks the Italian Space Agency (ASI) for
financial support.

The referee, Thierry Courvoisier, is thanked for his constructive
suggestions for improvements to this paper.

\end{acknowledgements}

\end{document}